\documentclass[preprint,showpacs,preprintnumbers,amsmath,amssymb]{revtex4}

\usepackage{graphicx}
\usepackage{dcolumn}
\usepackage{bm}

\begin{document}
\title{\Large \bf Chaos around charged black hole with dipoles}
\author{Chen Ju-hua}
\altaffiliation{cjh822@sina.com}
\author{Wang Yongjiu}
\affiliation{Department of Physics and Institute of Physics,\\
Hunan Normal University, Changsha, 410081, Peoples Republic of
China}
\begin{abstract}
We investigated dynamics of the test particle in the gravitational
field of the charged black hole with dipoles in this paper. At
first we have studied  the gravitational potential, by the
numerical simulations, we found, for appropriate parameters, that
there are two different cases in the potential curve, one is a
well case with a stable critical point, and the other is three
wells case with three stable critical points and two unstable
critical points. As consequence, the chaotic motion will rise. We
have performed the evolution of the orbits of the test particle in
phase space, we found  that the orbits of the test particle
randomly oscillate without any periods, even  sensitively depend
on the initial conditions and parameters. By performing
Poincar\'{e} sections for different values of the parameters and
initial condition, we have found regular motion and chaotic
motion. By comparing these Poincar\'{e} sections, we further
conformed that the chaotic motion of the test particle mainly
origins from the dipoles of the black hole.

\end{abstract}

\maketitle

\newpage
\section{Introduction}
Chaotic behaviors were considered as a interesting phenomena by
many physicists since Lorenz found the deterministic non-periodic
flow\cite{lorenz}. In the last decades,  chaos is one of the most
important idea used to explain various nonlinear phenomena in
nature. After the research on the three-body problem by
Poincar\'{e}, many studies about chaos in celestial mechanics and
astrophysics have been done and also have found the important role
of chaos in the universe\cite{Moser}\cite{wisdom}. There are two
main lines of research in general relativity, on deals with
chaoticity associated with inhomogeneous cosmological models, for
example, Oliveira {\sl et
al}\cite{oliveira}\cite{Monerat}\cite{Tonini}\cite{Ozorio} have
studied the chaotic behavior in Bianchi IX model. The other
assumes a given metric and looks for chaotic behavior of geodesic
motion in this background. Although we know many features of chaos
in Newtonian dynamics, we don't know, so far, so much about those
in general relativity. Because the gravitational field around a
black hole is very strong and nonlinear, we expect to find a new
type of chaotic behavior in such strong gravitational field which
does not appear in Newtonian dynamics
\cite{Misner}\cite{Belinskii}\cite{Barrow}. Some authors
\cite{Soota}\cite{Contopoulos}\cite{Dettmann}\cite{Yustsever}\cite{Karas}
\cite{Varvoglis} \cite{Bombelli}\cite{Moeckel}\cite{Letelier}
found chaotic behavior of a test particle in relativistic system.
Letelier {\sl et al} \cite{Gueron}\cite{Vieira}\cite{Letelie}
investigated the chaos in black hole with halos. J.H. Chen and
Y.J.Wang \cite{Chen} investigated the dynamics of a extreme
charged black hole; A. Saa \cite{Saa1}\cite{Saa2} extended
investigation on the integrability of oblique orbits of test
particle under the gravitational field corresponding to the
superposition of an infinitesimally thin disk and a monopole to
the more realistic case, for astrophysical purpose, of a thick
disk. And there are many examples in the literature of chaotic
motion involving black holes, in the fixed two centers problem
\cite{Contopoulos}\cite{Contopoulos1}\cite{Cornish}, in a black
hole surround by gravitational waves\cite{Bombelli1},
\cite{Leteli}, and in several core-shell models with relevance to
the description of galaxies\cite{Vieira1}. As to Newtonian case,
the recent works of C. Chicone {\sl et al} \cite{Chicone} on the
chaotic behavior of the Hill system. In order to investigate the
chaotic behavior of the dynamics system, Many researchers
concentrated on the study of chaotic dynamics in general
relativity with the Poincar\'{e}-Melnikov method
\cite{Melnikov}\cite{Wiggins}.The Melnikov method is an analytical
criterion to determine the occurrence of chaos in integrable
systems in which homoclinic (or heteroclinic) manifolds
biasymptotic to unstable critical points or to periodic orbits
(more generally to invariant tori) are subjected to small
perturbations.\\ It's well known that the charge
non-spheral-symmetry distribution of the charged black hole is a
popular phenomena. so it is interested to investigate the motion
of the test particle in the space-time of the charged black hole
with dipoles.  In this paper, by  performing  the numeral
simulations, we figured out orbital evolution in phase space and
Poincar\'{e} sections for different initial conditions.  we
confirm that there are regular and chaotic motion of the test
particle in the gravitational field of the charged black hole with
dipoles. The  others organize as follows: In the next section we
investigate the Hamiltonian of the charged black hole with
dipoles, In section III, we perform the numerical stimulations to
study the potential and the evolution of the orbits of the test
particle, at the same time we present Poinccar\'{e} sections for
different parameters. In the last section, a brief conclusion is
given.
\section{Hamiltonian of the charged black hole with dipoles}
Wang  {\sl et al}\cite{wang1}\cite{wang2}\cite{wang3}gave out the
metric of the charged black hole with dipoles. because the
space-time of our system is static axial-symmetry, by using
cylindrical $(r, \theta, z )$ to describe the space-time, we
obtain the potential which describe the gravitational field of
charged black hole with dipoles
\begin{eqnarray}
V=-\frac{M}{\sqrt{r^{2}+z^{2}}}+\frac{Q^2}{2(r^{2}+z^{2})}+
\frac{P^{2}z^{2}}{2(r^{2}+z^{2})^{3}},
\end{eqnarray}
where $M, Q, P $ are mass, charges and dipoles of the black hole,
respectively, and $r^{2}+z^{2}=x^{2}+y^{2}+z^{2}$, where $x, y, z$
are the usual Cartesian coordinates.\\
we know that the angular momentum $L$ in $z$ direction is
conserved and we can also easily reduce the three-dimensional
original problem to a two-dimensional one in the coordinates $(r,
z)$. Now we consider the bounded orbit for the test particle under
the potential (1), so the Hamiltonian is
\begin{eqnarray}
H=\frac{\dot{r}^{2}+\dot{z}^{2}}{2}+\frac{L^{2}}{2r^{2}}-\frac{M}{\sqrt{r^{2}
+z^{2}}}+\frac{Q^2}{2(r^{2}+z^{2})}+\frac{P^{2}z^{2}}{2(r^{2}+z^{2})^{3}}.
\end{eqnarray}
The Hamiltonian (2) is smooth everywhere, the corresponding
Hamiltonian-Jacobi equations can be properly separated in
parabolic coordinates \cite{Dorizzi}\cite{Grammaticos}, leading to
the second constant of the motion
\begin{eqnarray}
C=R_{z}-\alpha\frac{r^{2}}{2},
\end{eqnarray}
where $R_{z}$ is the $z$ component of the Laplace-Runge-Lenz
vector
\begin{eqnarray}
R=\frac{M}{\sqrt{r^{2}+z^{2}}}(r\hat{r}+z\hat{z})+V\times L,
\end{eqnarray}
where $L$ stands for the total angular momentum. In this case,
with two constants of motion $H$ and $C$, the equations for the
trajectories of the test particle can be reduced to quadrature in
parabolic coordinates \cite{Dorizzi}\cite{Grammaticos} and we
notice that the equation of motion are invariant under the
following rescaling:
\begin{eqnarray}
r\rightarrow\lambda r, z\rightarrow\lambda z, t\rightarrow\lambda
t, M \rightarrow\lambda M,\nonumber \\  Q\rightarrow\lambda Q,
P\rightarrow\lambda P, H\rightarrow\lambda H.
\end{eqnarray}
\section{Numerical simulations}
The gravitational potential (1) changes from one well to three
wells when we change one parameter while fixed the other
parameters. Fig.1 shows the potential for different values of $P$
($P^{2}=7 (dashed)$  and  $P^{2}=2$)(real) with fixed other
parameters ($M=1, Q^{2}=0.2, r=1$,). From Fig.1 we can see that
the motion in one well is oscillation, however in the three wells
potential, the motion is very different from the case of one well
case, there are three stable critical points (A) and two unstable
critical points (B), when the kinetic energy of the test particle
is much higher than the gravitational potential, the motion is
similar to the one well case, but when the kinetic energy of the
test particle is closed to the gravitational potential, the motion
oscillate randomly in one of the three wells, particularly near
the two unstable critical points, so the motion of the test
particle in this case sensitively depending on the initial
conditions and the parameters so that it becomes chaotic. The
Melnikov method is useful to find the regions of chaotic
oscillation. In this paper,  we will not use the analytical
method,we will figure out the evolution of the test particle in
phase space and its Poinccar\'{e} section method to investigate
the properties of the dynamics of the test particle in the
gravitational field of charged black hole with dipoles.
\cite{Abdullaev1}\cite{Abdullaev2}.\\
In order to investigate the evolutions of the test particle in the
gravitational field of the charged black hole with dipoles and its
chaotic behavior, we use variables $(\dot{r},\dot{z},r,z)$ and the
package POINCAR\'{E}\cite{Ceb-Terrab} to perform following
numerical experiments. Figs.2-5 show the evolution of the particle
in the compact phase space. We can see that the particle
oscillates randomly in the phase space with no periodic and
sensitively depends on the initial conditions and the parameters
of the system, that's to say, the properties of the motion is
chaotic if we choose properly parameters, which is expected by
studying the dynamics of the particle in the gravitational
potential of the charged black hole with dipoles.\\
To determine the chaotic behavior of a dynamical system, we can
further perform the Poincar\'{e} section in the phase space. If
the motion is not chaotic, the plotted points form a closed curve
in the two-dimensional $(\dot{r}, r)$ plane, because a regular
orbit will move on a torus in the phase space and the curve is a
cross section of the torus; If the orbit is chaotic, some of those
tori will be broken and the Poincar\'{e} section does not consist
of a set of closed curves, but the points will be distributed
randomly in the allowed region to form a chaotic  sea. From the
distribution of the points in Poincar\'{e} section, we can judge
whether or not the motion is chaotic. Figs. 6-11 show the typical
Poincar\'{e} sections across the plane $z=0$ for different initial
conditions. In Fig.6, we present the Poincar\'{e} section for
$M=1, L^{2}=0.4, P^{2}=12$, $Q^{2}=2$ and $H=-0.18800554$, the
section consists of KAM tori structure which characterizes a
quasi-integrable Hamilton system with quasi-periodic orbits. Under
these values of the parameters, there is no chaotic behavior. Then
in Fig.7, we plot the Poincar\'{e} sections for $M=1, L^{2}=0.4,
P^{2}=1.2$, $Q^{2}=0.2$ and $H=-0.28340554$, we can see that some
points have distributed randomly in a finite region to form a
chaotic sea and there two islands which are surrounded by the
chaotic sea. If we go on changing the parameters and the initial
conditions to draw Fig.8 for $M=1, L^{2}=0.4, P^{2}=12$,
$Q^{2}=0.2$, $H=-0.27800554$ and $M=1, L^{2}=0.2, P^{2}=12$,
$Q^{2}=0.2$, $H=-0.28911665$, further larger chaotic sea will
obtain in an allowed region and the islands will almost submerge
by the chaotic sea, this means that the quasi-periodic orbits
change into chaos. By comparing the parameters of Figs.6-9, we can
find that chaotic behavior mainly origins from the dipoles of the
black hole i.e.the non-sphere-symmetry of the charge distribution
in the black hole. \\
Performing further investigation, we will figure out the two
following extreme cases, Fig.10 present Poincar\'{e} section for
$M=1, L^{2}=0.4, P^{2}=1.2$, $Q^{2}=0$ and $H=-0.29340554$. In
this case the black hole is neutron, due to the un-sphere-symmetry
of the charge distribution, there are dipoles $P$ in the black
hole. From Fig.10, we can find that there are two islands surround
by a large chaotic sea, that's to say, under above parameters our
system is chaotic. However if the charges distribute
spheral-symmetry in the black hole, there is no dipole , so we
figure out Poincar\'{e} section for $M=1, L^{2}=0.4, P^{2}=0$,
$Q^{2}=0.2$ and $H=-0.28400554$. From Fig.11, we obviously see
that there is KAM tori structure in the section, this case
corresponds to an integrable motion.
\section{Conclusions and discussions}
In this paper we investigated the gravitational potential (1), by
the numerical stimulations, we find, for appropriate parameters of
the our system, that there are two cases in the potential curves,
one is a well case with a stable critical point, and the other is
three wells case with three stable critical points and two
unstable critical points. As consequence, the chaotic motion will
rise. In order to verify the chaotic motion, we have performed the
evolution of the orbits of the test particle in  phase space, we
can obviously see  that the orbits randomly oscillate without any
periods, even  sensitively depend on the initial conditions and
parameters. By the Poincar\'{e} section method, we have presented
Poincar\'{e} sections for different values of the parameters and
initial conditions. In these Poincar\'{e} sections we have found
regular motion and chaotic motion. By comparing these Poincar\'{e}
sections, we further conform that the chaotic motion of the test
particle mainly origins from the dipoles of the black hole.
\section{Acknowledgments}
Authors are very grateful for the helps of Professor Alberto Saa
(Departamento de Matem$\acute{a}$tica Aplicada, IMECC-UNICANP
Campinas. S.P. Brazil ), Professor H. P. de Oliveira
(NASA/Fermilab Astrophysics Center, Fermi National Accelaratory
Batavia, $\Pi$linois) and Professor Wenhua Hai, Professor Jiliang
Jing, Professor Hongwei Yu (Department of Physics, Hunan Normal
University, People Republic of China).
\newpage

\newpage

\begin{figure}
\includegraphics{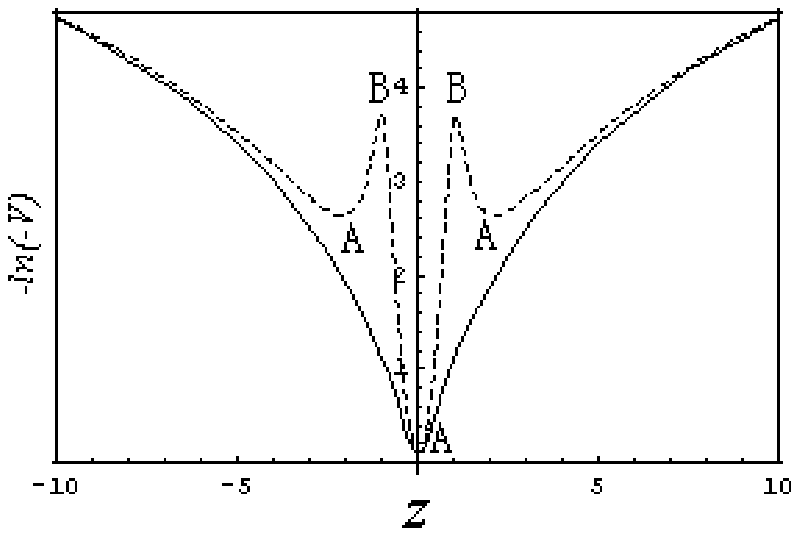}
\caption{Plots of Gravitational potential (1) for $M^{2}=1,
Q^{2}=0.2, r=1$ and $P^{2}=7$ (dashed), $2$ (real). From the plots
we can see that when $P^{2}=2$ (real) there only one well in the
potential curve and a stable critical point (A); but when
$P^{2}=7$ (dashed) there are three wells in the potential curve
and three stable critical points (A) and two unstable critical
points (B). the test particle oscillates randomly in one of three
wells.}
\end{figure}
\begin{figure}[htbp]
\begin{center}
\includegraphics{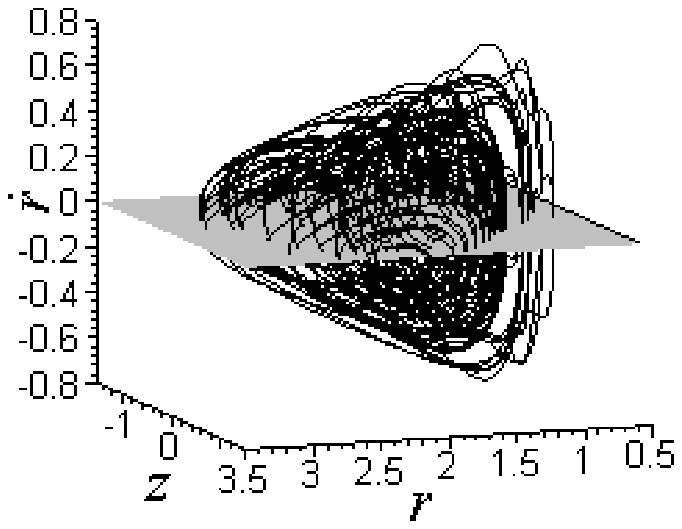}
\end{center}
\caption{3Dimesional(\.{r}, r, z) plot of the evolution of the
orbit of the test particle in the charged black hole with dipoles
in the phase space for $M=1, L^{2}=0.4, P^{2}=12$, $Q^{2}=0.2$ and
$H=-0.27800554$.}
\end{figure}
\begin{figure}[htbp]
\begin{center}
\includegraphics{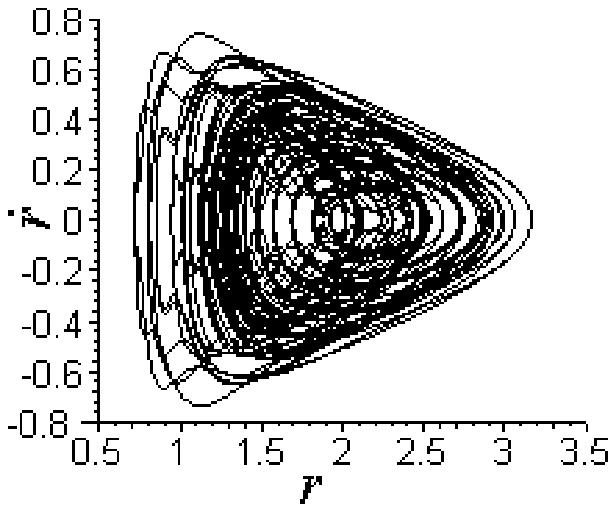}
\caption{2Dimesional(\.{r}, r) plot of the evolution of the orbit
of the test particle in the charged black hole with dipoles in the
phase space for $M=1, L^{2}=0.4, P^{2}=12$, $Q^{2}=0.2$ and
$H=-0.27800554$.}
\end{center}
\end{figure}
\begin{figure}[htbp]
\begin{center}
\includegraphics{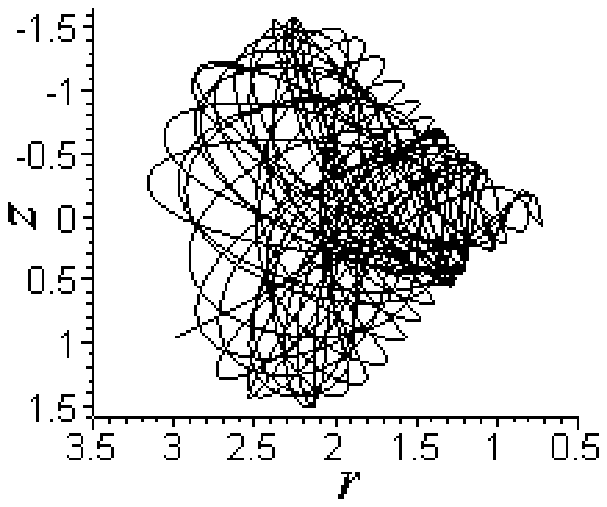}
\end{center}
\caption{2Dimesional(r, z) plot of the evolution of the orbit of
the test particle in the charged black hole with dipoles in the
phase space for $M=1, L^{2}=0.4, P^{2}=12$, $Q^{2}=0.2$ and
$H=-0.27800554$.}
\end{figure}
\begin{figure}[htbp]
\begin{center}
\includegraphics{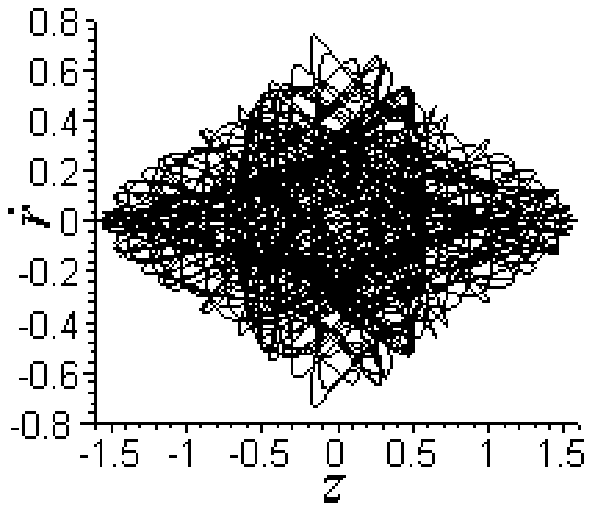}
\end{center}
\caption{2Dimesional( \.{r}, z) plot of the evolution of the orbit
of the test particle in the charged black hole with dipoles in the
phase space for $M=1, L^{2}=0.4, P^{2}=12$, $Q^{2}=0.2$ and
$H=-0.27800554$.}
\end{figure}
\begin{figure}[htbp]
\begin{center}
\includegraphics{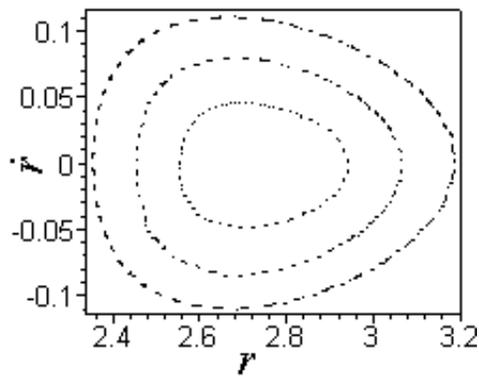}
\end{center}
\caption{Pioncar\'{e} section (\.{r}, r) across the plane $z=0$,
\.{z} $>$ 0 for $M=1, L^{2}=0.4, P^{2}=12$, $Q^{2}=2$ and
$H=-0.18800554$. For these values of the parameters we have the
section of an integrable motion. }
\end{figure}
\begin{figure}[htbp]
\begin{center}
\includegraphics{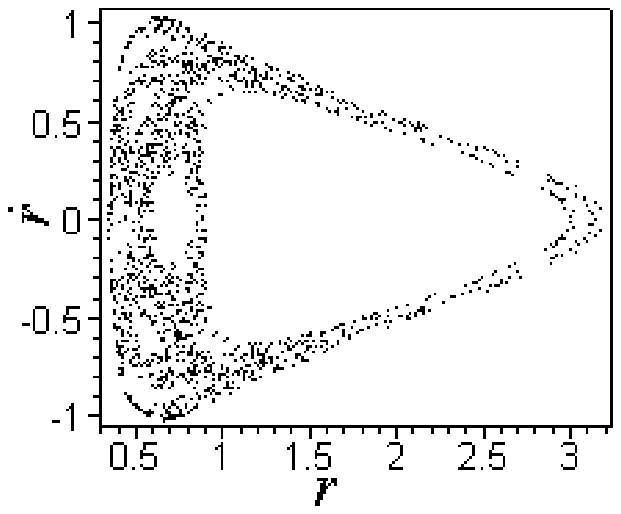}
\end{center}
\caption{Pioncar\'{e} section (\.{r}, r) across the plane $z=0$,
\.{z} $>$ 0 for $M=1, L^{2}=0.4, P^{2}=1.2$, $Q^{2}=0.2$ and
$H=-0.28340554$. Under these conditions there is a large chaotic
sea, which means chaotic motion of the test particle, in the
section.}
\end{figure}
\begin{figure}[htbp]
\begin{center}
\includegraphics{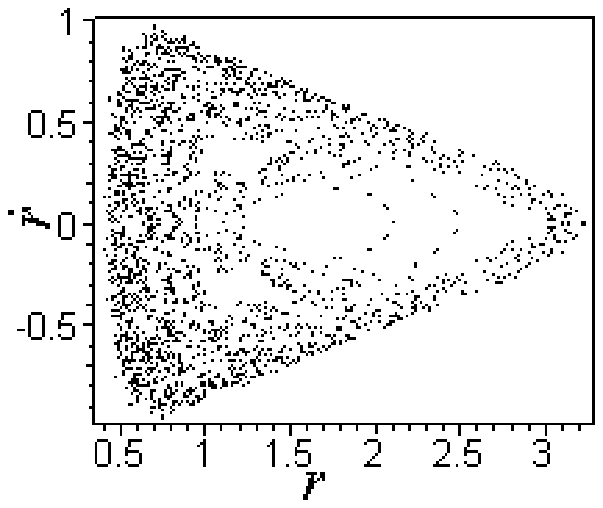}
\end{center}
\caption{Pioncar\'{e} section (\.{r}, r) across the plane $z=0$,
\.{z} $>$ 0 for $M=1, L^{2}=0.4, P^{2}=12$, $Q^{2}=0.2$ and
$H=-0.27800554$.¡¡Under these conditions there is a larger chaotic
sea, which means chaotic motion of the test particle, than the
precedent figure in the section. }
\end{figure}
\begin{figure}[htbp]
\begin{center}
\includegraphics{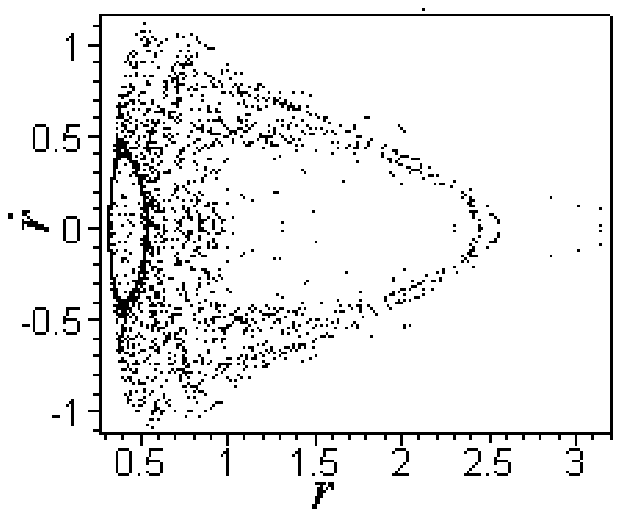}
\end{center}
\caption{Pioncar\'{e} section (\.{r}, r) across the plane $z=0$,
\.{z} $>$ 0 for $M=1, L^{2}=0.2, P^{2}=12$, $Q^{2}=0.2$ and
$H=-0.28911665$. Under these conditions there is a much larger
chaotic sea, which means chaotic motion of the test particle, than
the precedent two figures in the section.}
\end{figure}
\begin{figure}[htbp]
\begin{center}
\includegraphics{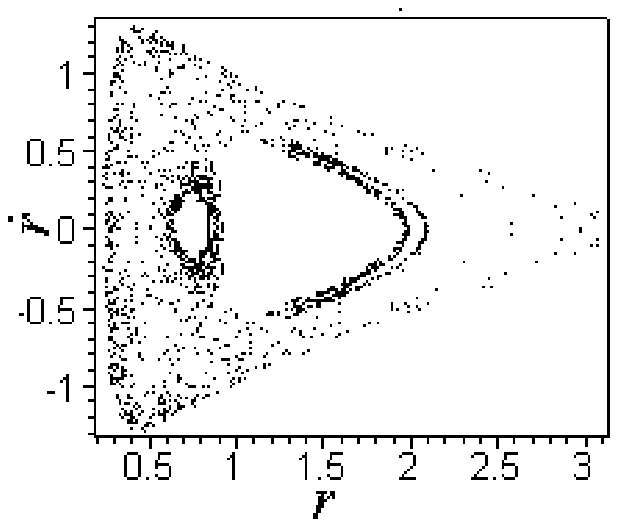}
\end{center}
\caption{Pioncar\'{e} section (\.{r}, r)across the plane $z=0$,
\.{z} $>$ 0 for $M=1, L^{2}=0.4, P^{2}=1.2$, $Q^{2}=0$ and
$H=-0.29340554$. Under these extreme conditions there is large
chaotic sea, which means chaotic motion of the test particle, in
the section.}
\end{figure}
\begin{figure}[htbp]
\begin{center}
\includegraphics{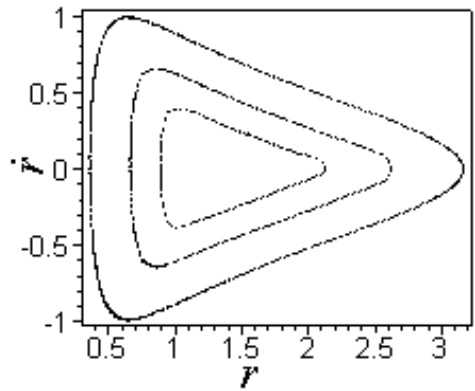}
\end{center}
\caption{Pioncar\'{e} section (\.{r}, r)across the plane $z=0$,
\.{z} $>$ 0 for $M=1, L^{2}=0.4, P^{2}=0$, $Q^{2}=0.2$ and
$H=-0.28400554$.  For these extreme  values of the parameters we
have the section of an integrable motion just as Fig.6.}
\end{figure}
\end{document}